\documentclass[12pt,preprint]{aastex}

\def\degree{\ifmmode {^\circ}\else {$^\circ$}\fi}
\def\rstar{\ifmmode {\, R_{\star}}\else $R_{\star}$\fi}
\def\msol{\ifmmode {\, M_{\odot}}\else $M_{\odot}$\fi}
\def\rsol{\ifmmode {\, R_{\odot}}\else $R_{\odot}$\fi}
\def\lsol{\ifmmode {\, L_{\odot}}\else $L_{\odot}$\fi}
\def\msolyr{\ifmmode {\, M_{\odot}\,{\rm yr}^{-1}}\else $M_{\odot}\,{\rm yr}^{-1}$\fi}
\def\mdot{\ifmmode {\,\dot{M}}\else $\dot{M}$\fi}
\def\mdotyr{\ifmmode {\,\dot{M}\,yr^{-1}}\else $\dot{M}\,yr^{-1}$\fi}

\begin{document}

\title{An Extremely Bright Echo Associated With SN 2002\lowercase{hh}}

\author {D.L. Welch\altaffilmark{1}, Geoffrey C. Clayton\altaffilmark{2,3}, Amy Campbell\altaffilmark{2}, M.J. Barlow\altaffilmark{4}, Ben E.K. Sugerman\altaffilmark{5,6}, Margaret Meixner\altaffilmark{5}, and S.H.R. Bank\altaffilmark{3,7}}

\altaffiltext{1}{Dept. of Physics \& Astronomy, McMaster University, Hamilton, Ontario,  L8S 4M1 Canada; welch@physics.mcmaster.ca}
\altaffiltext{2}{Dept.~of Physics \& Astronomy, Louisiana State
University, Baton Rouge, LA 70803;
gclayton@fenway.phys.lsu.edu, campbell@theory.phys.lsu.edu}
\altaffiltext{3}{Maria Mitchell Observatory, 4 Vestal St., Nantucket, MA 02554}
\altaffiltext{4}{Dept. of Physics and Astronomy,
	University College London, Gower Street, London WC1E 6BT, UK; mjb@star.ucl.ac.uk}
\altaffiltext{5}{Space Telescope Science Institute, 3700 San Martin Dr., Baltimore, MD 21218; meixner@stsci.edu}
\altaffiltext{6}{Goucher College, Dept of Physics, 1021 Dulaney Valley Rd.,
Baltimore, MD 21204; ben.sugerman@goucher.edu}
\altaffiltext{7}{Department of Physics, UMBC, 1000 Hilltop Circle, Baltimore, MD 21250; sbank1@umbc.edu}

\begin{abstract}

We present new, very late-time optical photometry and spectroscopy of 
the interesting Type II-P supernova, SN 2002hh, in NGC 6946. Gemini/GMOS-N 
has been used to acquire visible spectra at six epochs between 2004 August 
and 2006 July, following the evolution of the SN from age 661 to 1358 days.
Few optical spectra of Type II supernovae with ages greater than one year 
exist. In addition, $g'r'i'$ images were acquired at all six epochs. 
The spectral and photometric evolution of SN 2002hh has been very unusual. 
Measures of the brightness of this SN, both in the R and I bands as well as 
in the H$\alpha$ emission flux, show no significant fading over an 
interval of nearly two years. The most straightforward explanation 
for this behavior is that the light being measured comes not only 
from the SN itself but also from an echo off of nearby dust. Echoes 
have been detected previously around several SNe but these echoes, 
at their brightest, were $\sim$8 mag below the maximum brightness of 
the SN. At V$\sim$21 mag, the putative echo dominates the light of 
SN 2002hh and is only $\sim$4 mag below the outburst's peak brightness. 
There is an estimated 6 magnitudes of total extinction in V towards SN 
2002hh. The proposed explanation of a differential echo/SN absorption 
is inconsistent with the observed BVRI colors.
\end{abstract}



\section{Introduction}

The condensation of dust in supernova (SN) ejecta is not well constrained by observations
since the detection of newly-formed dust in young SNe in nearby galaxies 
has been extremely rare. Signatures of condensing dust can be observed 
in nearby Type II SNe approximately 1-2 years after core collapse. There are 
three strong indications of the formation of dust:
1) a sudden decrease in continuum brightness in the visible due to increased 
dust extinction, 2) an accompanying development of an infrared excess 
arising from dust grains absorbing higher-energy photons and re-emitting 
them in the infrared, and 3) the development of asymmetric, blue-shifted 
emission-line profiles \citep[e.g.,][]{2006Sci...313..196S}. This
combination of observable effects can be best explained by dust forming 
within the expanding ejecta and preferentially extinguishing emission from 
the far-side (hence red-shifted) gas which also results in a shift of the
centroid of the emission line profiles to the blue, optical 
dimming, and mid-infrared brightening \citep{1989LNP...350..164L, 1991supe.conf...82L}. Asymmetric emission line profiles have been seen in SNe 1987A, 1998S, 
1999em, 2003gd and possibly 2004et  \citep[e.g.,][]{1991supe.conf...69D,2000AJ....119.2968G,2003MNRAS.338..939E,2004MNRAS.352..457P,2006Sci...313..196S, 2006astro.ph..8432S}.
In the sample of seven supernovae with such spectra, dust 
appears to have formed in the ejecta of at least five. Even within 
this small sample the dust formation times range from near 300 days to 
550 days. 

SN 2002hh was discovered in NGC 6946 on 2002 October 31 \citep{2002IAUC.8005....1L}. With the subsequent discovery of SN 2004et, there have now been eight SNe discovered in NGC 6946 in less than a century. 
A spectrum of SN 2002hh, obtained on 2002 November 2, showed a broad but very low 
contrast H$\alpha$ emission and absorption features with a very red, 
nearly featureless continuum indicating a very young Type II SN 
\citep{2002IAUC.8007....2F}. Optical imaging of the region shows extensive
dust obscuration in this region of NGC 6946 and \citet{2002IAUC.8024....1M} 
estimate a total absorption of A$_V\sim$6.1 mag toward SN 2002hh of which 
about 1 mag is Galactic foreground obscuration. This estimate was made by 
comparing the observed J-K$_s$ color to template IR lightcurves 
\citep{2001MNRAS.324..325M} and is supported by the strong interstellar 
Na I D absorption seen in the spectrum \citep{2002IAUC.8007....2F}.
CO emission was seen in K-band spectra of SN 2002hh in 
the first 6 months after outburst \citep{2004MNRAS.352..457P}.
The lightcurve of SN 2002hh shows it to be a Type II-P, where the SN remains at or near maximum brightness for $\sim$3 months after the explosion \citep{2006MNRAS.368.1169P}.

Thermal emission from dust in or near SN 2002hh was detected at mid-infrared 
wavelengths from 590 to 994 days by the {\it Spitzer Space Telescope} (SST) and confirmed by 
higher angular resolution Gemini/Michelle observations 
\citep{2005ApJ...627L.113B, 2006ApJ...649..332M}. An optically-thick dust shell has been inferred
having a mass of 0.04-0.15 M$_{\sun}$ and suggesting a massive M supergiant 
or luminous blue variable precursor.
The inner radius of the shell is estimated to be $\sim$1 x 10$^{17}$ to 
$\sim$1 x 10$^{18}$ cm  \citep{2005ApJ...627L.113B,2006ApJ...649..332M}.
The highest velocity gas from the 
SN 2002hh explosion will not reach this circumstellar dust for approximately 
28 yr \citep{2006ApJ...649..332M}. \citet{2005ApJ...627L.113B} could not rule 
out the formation of new dust in the ejecta of SN 2002hh, since any
mid-infrared emission from new dust would be swamped by emission from the 
existing circumstellar dust which would occupy a far larger solid angle. 

\citet{2006MNRAS.368.1169P} first suggested the existence of an IR echo 
in the JHK bands around SN 2002hh at early epochs up to 314 days. 
\citet{2006ApJ...649..332M} found that the mid-IR flux declined by about 
10\% from 590 to 994 days. They suggested that most of the mid-IR flux 
may have come from dust in the star formation region associated with 
the SN 2002hh precursor. They found no evidence for new dust condensing 
in the ejecta nor did they attribute the bulk of the emission to heated 
circumstellar material. The small declining component of the IR flux is 
attributed to an IR echo. An echo around SN 2002hh has been confirmed in the 
optical by Sugerman et al. (2007, in preparation), who have found a 
small but well-resolved circular scattered-light echo in visible light 
roughly 0\farcs18 in diameter at the astrometric position of the supernova. 

In this paper we present a series of new, late-time optical spectra and 
photometry of the Type II-P SN 2002hh which were obtained to look for 
evidence of dust formation in its ejecta.

\section{Observations}

Spectroscopy was obtained using the Gemini North telescope 
and the GMOS-N instrument during semesters 2004B, 2005B, and 2006A. 
Three spectra of duration 900 s each were obtained on 2004 August 20. 
These were obtained in longslit mode with a slit orientation of 
190\degree~in order to make use of a bright guide star. A 0\farcs75 
slit width was used for the observations with the B600-G5303 grating 
in first-order. The central wavelengths of the spectra were 5720, 5700, 
and 5680 \AA, respectively, to ensure that the two gaps between GMOS-N's 
three CCD chips could be bridged. A 2x2 binning of the CCD pixels in 
the low gain setting was employed.

Two additional epochs for SN 2002hh in 2004 were obtained using Director's 
Discretionary time. On 2004 October 8, three spectra, each of 720 s duration,
were obtained with a setup identical to that described above. The 
observations had central wavelengths of 5970, 5950, and 5930 \AA, 
respectively. These longer central wavelengths were selected in order 
that the  [Ca II] $\lambda\lambda$7291, 7324 lines would be recorded. 
A second set of spectra were obtained on 2004 December 11 UT.  
Each had a duration of 720 s and had central wavelengths of 5970, 5950, 
and 5930 \AA, respectively. 

Two epochs of spectra were obtained in 2005. The configuration of GMOS-N 
was identical to that above. The first epoch was obtained on 2005 August 4. 
Each exposure had a duration of 900 s to reduce the effect of fading of the 
SN on the signal-to-noise of the spectra, and had central wavelengths of 
5970, 5950, and 5990 \AA, respectively.  The second epoch was obtained on
2005 October 4 and each spectrum then also had an exposure of 900 s. 
The central wavelengths were 5970, 5950, and 5990 \AA, respectively. 

The final epoch of spectra reported in this paper were obtained under an 
identical GMOS-N configuration on 2006 July 18. Each had a duration of 
900 s and had central wavelengths of 5970, 5950, and 5990 \AA, respectively. 

All spectra had adjacent ``GCal" flat  exposures taken prior
to changes to new central wavelengths. CuAr spectra used to
calibrate the pixel-to-wavelength transformation were obtained
during daytime. A single set of bias frames were used to reduce
all of the spectra. The spectra from each individual night were 
averaged and normalized. 

Spectra were reduced using IRAF version 2.12.2a and version 1.8
of the IRAF external package {\sl gemini}. The usual pattern of
reduction tasks {\sl gbias}, {\sl gsflat}, {\sl gsreduce}, 
{\sl gswavelength}, {\sl gstransform}, {\sl gsskysub}, and 
{\sl gsextract} was employed. Sky subtraction regions were identified
by eye and usually incorporated about 50 pixels on either side of the
SN spectrum itself. Observations were not fluxed due to the
lack of photometric conditions on several of the nights and also due
to the primary goal of recording line profile variations. The radial 
velocity of NGC 6946 is $+$48 km s$^{-1}$. Emission lines from 
incompletely-subtracted HII regions in the SN 2002hh spectra allow 
us to register the supernova to the local standard of rest in the galaxy.
The spectra are shown in Figure 1 and are summarized in Table 1. (The spectra
in Figure 1 have not been de-reddened.) We adopt 2002 October 29 as the 
explosion epoch \citep{2006MNRAS.368.1169P}.

The position of the peak of the H$\alpha$ emission line is poorly determined
due to the contamination of the $\lambda\lambda$6548-6584 [NII] doublet and 
$\lambda$6563 H$\alpha$ line from the HII region. To gauge the
change in line position and symmetry, we instead measured the wavelength at
which the ascending and descending portions of the line profile reached
50\% of their peak values. Prior to making this measurement, a linear
continuum slope correction was applied based on the average intensity 
levels in the $\lambda\lambda$6240-6260 and 6940-6960 regions of the spectra.
The intensity of the H$\alpha$ line changes rapidly through the 50\%
points, ensuring precise estimates for the ascending and descending
half-peak wavelengths. In Table 1 we list, from left to right, the 
UT observation date, the age of the supernova, the wavelength where 
the ascending portion of the H$\alpha$ line reached 50\% of its peak 
value, and the corresponding position on the descending wing. The 
uncertainty in these wavelengths is $\pm$2\AA\ based on the repeatability 
of the number for each of the three or four spectra obtained on a given 
observation date.

On each of the six epochs described above, SN 2002hh was also observed in 
imaging mode with GMOS-N in the $g'$,$r'$, and $i'$ bands. Exposure times 
were 60 s. The instrumental $g'$,$r'$, and $i'$ magnitudes were converted 
to standard Johnson-Cousins $R$ and $I$ magnitudes using the photometric 
sequence from Table 1 of \citet{2006MNRAS.368.1169P}. The standard 
stars which fell on the central chip of GMOS-N (the chip on which SN 2002hh 
was imaged) were used for this purpose.

A simultaneous least-squares solution involving common color terms
but floating zeropoints for each night was used to transform the
instrumental $g'$,$r'$, and $i'$ ALLFRAME magnitudes to the $V$, $R$,
and $I$ sequence from \citet{2006MNRAS.368.1169P}. The uncertainties
in the  $V$, $R$, and $I$ transformations thus derived were 0.033, 0.014,
and 0.010 mags, respectively. All transformations were confirmed to be
linear with the bulk of the higher uncertainty in the $V$ transformation
being due to the residuals of two stars of intermediate color, 13 and 14.

The transformations were then inverted to allow the estimation
of the  $V$, $R$, and $I$ magnitudes from the instrumental $g'$,$r'$, and $i'$
magnitudes. The derived transformations are,\\
\noindent
$V = 0.7111 \times g' -0.0829 \times r' +0.3728 \times i' + zero~point$\\
$R = 0.0000 \times g' +0.7540 \times r  +0.2264 \times i' + zero~point$\\
$I = 0.0000 \times g' -0.2786 \times r' +1.2531 \times i' + zero~point$.\\

We were not able to provide V magnitudes for SN 2002hh because the supernova
light was too confused with other nearby sources of similar brightness in
every image taken in the $g'$ bandpass.  On 2006 July 18, SN 2002hh could 
not be measured accurately in $r'$ band image for similar reasons, but if it 
is assumed that the color of the SN was similar to that of the Pozzo standard 
number 5, then, I = 19.60 mag. (The small coefficient of the r' magnitude in
the transformation to I makes the transformation reasonably robust to
reasonable values for the color.) The measured R and I magnitudes of 
SN 2002hh for the six epochs are listed in Table 1 and plotted in Figure 2. 

\section{Discussion}

\subsection{Spectra}

\citet{2006MNRAS.368.1169P} observed SN 2002hh in the optical and near-IR 
from day 3 to 397.
Optical spectra, including H$\alpha$ and [O I] $\lambda\lambda$6300, 6364, 
were obtained at seven epochs. Pozzo et al.  commented 
that, in general, the spectra of SN 2002hh resemble those of SN 1987A at 
similar epochs. In particular, SN 1987A and SN 2002hh both showed P-Cygni 
profiles in H$\alpha$ and Na I D at early times \citep{1991supe.conf...36P}. 
The biggest difference between the SN 2002hh and SN 1987A
was the strength of the [Ca II] $\lambda\lambda$7291, 7323 and the Ca II 
near-infrared triplet which was much stronger in SN 1987A \citep{2006MNRAS.368.1169P}.
Their last spectrum, obtained on day 397, shows no evidence for a developing asymmetry due to dust forming in the ejecta. 

Figure 1 shows the new spectra obtained at an additional six epochs which
extend the spectral coverage to day 1358. None of the new spectra reveal
an asymmetry in H$\alpha$ similar to that which developed in SN 1987A 
around day 526, and which is believed to have indicated 
dust formation. (See Figure 7 of \citet{1991supe.conf...36P}.) The H$\alpha$ 
profile in SN 2002hh never completely lost its P-Cygni shape, having a 
broader red wing than a blue wing  \citep{1995A&A...297..802D}. 
Not only is there no development of a blue asymmetry but the 
FWHM of H$\alpha$ has increased in the day 661-1358 spectra compared to 
the last epoch of \citet{2006MNRAS.368.1169P} at day 397 without any 
statistically-significant shift in line centroid. No increase in the 
opacity in the red wing 
of the line is seen in SN 2002hh. In fact, the red and blue wings both become 
broader between day 397 and 661. No such 
broadening of H$\alpha$ was seen in SN 1987A \citep{1991supe.conf...36P}. 
 As seen in Table 1, the half-power points of H$\alpha$ show no
sensible evolution of either position over the course of the observations
reported here. 
 
The fact that the H$\alpha$ emission line profile has not developed a blue asymmetry but rather has retained its P-Cygni shape and broadened since day 397 may be due to the presence of an echo around SN 2002hh. 
It is possible that dust formation in 
2002hh is perhaps hidden or its interpretation confounded by scattered light 
from an echo resulting in
a systematic error in the estimated age of the spectral features in the
optical  \citep{2006ApJ...649..332M}. 
However,  \citet{2006MNRAS.368.1169P} note that the spectrum of SN 2002hh obtained on day 397 has developed a strong blue continuum indicating that the echo may have been affecting the light from SN 2002hh even at this early epoch. 

Since SN 2002hh is a Type II-P, it stayed near maximum brightness for $\sim$3 
months. Although there is a gap in the photometry of SN 2002hh from day 43 
to 166, its lightcurve behavior was consistent with both SN 1987A and 
SN 1999em \citep{2006MNRAS.368.1169P}. Any SN light echo pulse width will be 
quite broad since it will contain an integral of the SN light over the first 
three months. \citet{2006PASP..118..351V} estimated a pulse width of 
$\sim$43 d in the B band and $\sim$138 d in the R band for for the 
Type II-P SN 2003gd. The pulse width of SN 2002hh is likely to be on the 
same order of the length of the plateau phase of the lightcurve, $\sim$100 d. 
The echo spectrum seen at late times, would have been an average of the SN 
spectrum over this period. \citet{2006MNRAS.368.1169P} reported spectra from 
SN 2002hh at days 4, 8, and 44. At day 44, the lines in the SN spectrum had 
already taken on the P-Cygni profile shape that was seen all the way out to 
day 1358.

The late-time Gemini/GMOS spectra do resemble the Pozzo et al. day 44 
spectrum of SN 2002hh. The day 44 spectrum and the new late-time 
(day 661-1358) spectra presented here are all broader than the last 
Pozzo et al. epoch at day 397. The H$\alpha$ emission profiles from the 
six Gemini/GMOS epochs spanning over 700 days are almost identical and 
show no significant profile evolution with time as seen in Figure 1. 
We find that the extrema of the H-alpha wings indicate an expansion
velocity between 11,000-13,000 km s$^{-1}$ for our observed epochs,
where the uncertainty is dominated by the assumption of the continuum
locations. The FWHM of all six epochs are $\sim$5600 km s$^{-1}$. This 
is significantly broader than the profiles from days 162 to 397 which 
have FWHM of 
$\sim$4500 km s$^{-1}$ \citep{2006MNRAS.368.1169P}. No spectra were taken
between days 44 and 162, but the day 44 spectrum was in the middle of the 
time period of the light pulse which we believe is being scattered toward 
us in the echo. These late-time H$\alpha$ profiles, broader than the 
previously observed epoch (day 397), support the interpretation that the 
late-time spectrum is dominated by the echo. 

To investigate the contribution from the echo, the counts from SN 2002hh 
in the H$\alpha$ emission line profiles was examined. In Figure 2, 
the peak counts in H$\alpha$ in the Gemini/GMOS spectra from all six epochs, 
are plotted together, scaled to their exposure times along with the epochs observed by \citet{2006MNRAS.368.1169P}. The spectra could not
be fluxed since no standards were observed but they were acquired with the 
same slit with the same setup and at similar airmasses during each observation.
The counts should be roughly representative of the relative fluxes taking 
into account the uncertainties ($\sim$30\%) introduced by differences in 
seeing and transparency. In particular, the lowest count rates were associated
with the worst seeing, which occurred on days 710 and 1358. This comparison
in Figure 2 shows that there is no evidence for significant fading of the
peak flux in H$\alpha$ over the period from day 661 to 1358. We also plot
the evolution of SN 1987A's scaled peak H$\alpha$ flux \citep[and the SUSPECT 
supernova spectrum database]{1993MNRAS.262..313C} in Figure 2.  
If SN 2002hh had faded in a manner similar to SN 1987A then
from day 397 \citep{2006MNRAS.368.1169P}  to day 1358, H$\alpha$ should have 
faded by a factor of $\sim$10$^4$.  However, SN 2002hh's H$\alpha$ flux only 
drops by a factor of 2 from 397 to 1358. Since SN 2002hh is a Type II SN, 
its massive precursor star likely was located in a star formation region 
with significant H$\alpha$ flux. We have decomposed the narrow H II region 
emission lines and compared their flux to the broad SN H$\alpha$ emission. 
On day 661, the narrow emission lines made up only 16\% of the flux. 
So most of the flux is from the SN, which is distinguished in any case by 
its very broad H$\alpha$ emission up to day 1358. The fact that the broad 
H$\alpha$ emission profile can be seen and measured even at day 1358, 
means that we are still measuring SN 2002hh without significant contamination 
from other sources. 

\subsection{Photometry}

The latest published V-band photometry of SN 2002hh, reported by  
\citet{2006MNRAS.368.1169P}, was on day 269 - too early to expect to see 
evidence of fading due to dust formation in the ejecta and also before the 
echo became noticeable in the optical. 
\citet{2006MNRAS.368.1169P} detected a $K-L'$ excess that developed in 
photometry of SN 2002hh between days 200 and 314. This may indicate the 
presence of an infrared echo, presumably produced by early-time heating of 
the circumstellar dust by the SN.
\citet{2006ApJ...649..332M} found no evidence for new dust condensing in 
the ejecta and suggested that most of the mid-infrared flux may have come from 
dust in the star formation region associated with SN 2002hh's precursor.
They also found that the mid-infrared flux declined by about 10\% from day 590 
to 994. The small declining component of the infrared flux was attributed 
to an infrared echo. 

The brightness of the echo at optical wavelengths was highly unexpected - the 
late-time observed brightness of SN 2002hh in the $V$ band was only 4 mag 
fainter than the lightcurve peak of $V=17.2$ mag 
\citep[Sugerman et al. 2007, in preparation]{2006MNRAS.368.1169P}.  
The photometry is also consistent with the lack of fading seen in the 
H$\alpha$ emission shown in Figure 2.  Typically, one would expect a light 
echo to appear fainter than the SN maximum by 8 mag or more 
\citep{2005MNRAS.357.1161P}. 

The fading of SN 2002hh, from its maximum brightness of R$\sim$15.5 mag
and I$\sim$14.3 mag leveled off around R$\sim$20 mag and I$\sim$19 mag 
sometime between day 269, the last \citet{2006MNRAS.368.1169P} data point, 
and day 661, the first data point obtained here. This behavior is very 
different from the SN 1987A lightcurve which is scaled and plotted for 
comparison in Figure 3. If SN 2002hh had behaved like SN 1987A, it would 
have been about 3-4 mag fainter in the $R$ and $I$ bands than observed on 
day 661. SN 2002hh seems to have faded slowly from $I\sim$19.0 mag to 
$I\sim$19.6 mag between days 661 to 1358. By day 1358, SN 2002hh was 
$\sim$9 mag too bright at $I$ when compared to SN 1987A. 

Observations of SN 2002hh in the optical were made with the 
High-Resolution Camera (HRC) of the Advanced Camera for Surveys 
(ACS) aboard {\em HST} on 2005 Sep 17, 2006 Apr 23, and 2006 Nov 18, 
respectively days 1054, 1272, and 1481 after maximum. As will be fully 
presented in Sugerman et al. (2007, in preparation), these observations 
reveal a small but well-resolved circular scattered-light echo 
roughly 0\farcs18 in diameter at the astrometric position of the 
supernova.  At an estimated distance of 5.9 Mpc, this implies the 
scattering dust is roughly 12 ly in front of the SN. The integrated 
fluxes in F435W, F606W, and F814W from the first epoch roughly 
convert to Johnson-Cousins $B$, $V$, and $I$ magnitudes of 
24.3, 21.4, and 19.5 mags, respectively \citep{2005PASP..117.1049S}. 
The $V$-band brightness decreased by only a few hundredths of a mag 
in the subsequent two epochs. These measurements are fully consistent with 
ours and show that SN 2002hh had still only faded by $\sim$4 mag at 
$V$ from its maximum brightness by day 1481. 

By day 661, a typical Type II-P SN with no echo will have faded $\sim$7.5 mag
below its maximum brightness. Normally, an optically-thin dust echo will 
get brighter as the optical depth of the dust increases 
\citep{2005MNRAS.357.1161P}. Patat's sheet model (R=200, $\Delta$R=50), 
assuming dust with $\tau$=0.03, will produce an echo with a brightness 
$\sim$11.75 mag below the SN maximum, while dust with $\tau$=2.5 will 
produce an echo with a brightness only $\sim$6.75 mag below the SN 
maximum. If multiple scattering is taken into account, as it must be 
when $\tau >$1, then $\tau$=2.5 will produce an echo with a brightness 
$\sim$9.25 mag below the SN maximum or three magnitudes fainter than 
in the single scattering model. 

There has been significant discussion in the literature regarding the 
nature of the dust along the direct line-of-sight to SN 2002hh. 
\citet{2006MNRAS.368.1169P} reconsidered the extinction for SN 2002hh in
some detail and concluded that a line-of-sight ``dust pocket'' containing
atypically small dust grains contributed to a total A$_V$ = 5.3 mag  as a
result of a two-component dust model (A$_V$ = 3.3 mag and R$_V$ = 3.1 
plus A$_V$ = 1.7 mag and R$_V$ = 1.1). It is, a priori, rather unlikely 
that there is dust in NGC 6946 with R$_V$ = 1.1. Another possible
explanation is that there is a mismatch between the intrinsic SED of 
SN 2002hh and that of SN 1999em which was used as a lightly-reddened 
template.  \citet{2005ApJ...635L..33W} alternatively suggested that light 
scattered from circumstellar dust near the SN will cause apparently small 
values of R$_V$. It should be noted that in Wang's model, the inner radius 
of the shell is at 10$^{16}$ cm whereas \citet{2006MNRAS.368.1169P} suggest 
that dust within 10$^{17}$ cm of the SN would be evaporated. 
The apparent low value for R$_V$ could also be caused by combining dust 
with two different values of R$_V$ \citep{2005ApJ...624..118M}. Nevertheless, 
the possibility of anomalous dust must be seriously considered when modeling 
the echo in SN 2002hh. 

In one possible scenario, the extreme brightness of 
the SN 2002hh echo can be explained if the optical depth is much greater 
($\sim$6 mag) along the line of sight to the SN than in other nearby
directions ($\lesssim$1 mag). The SN light from the echo could 
skirt the large dust optical depth sitting directly in front of 
SN 2002hh and be bright by comparison. See Figure 1 of 
\citet{2006MNRAS.369.1949P}. The observed lightcurve and spectral
evolution could be accounted for by the proposed differential total
absorption. 

Using the time-integrated spectrum of SN 1999em, reddened, as described 
by \citet{2006MNRAS.368.1169P}, to reproduce the expected light pulse 
from SN 2002hh, the predicted observed colors for the SN 2002hh echo 
are $B$-$V$ = 2.7, $V$-$I$=3.1 and $R$-$I$ = 1.3 mag. These are consistent 
with the colors actually observed for SN 2002hh at early epochs 
\citep{2006MNRAS.368.1169P}. However, at late epochs ($>$600 days) when 
the light of SN 2002hh was apparently dominated by the echo light, the
colors were, $B$-$V\sim$2.6, $V$-$I\sim$1.8 and $R$-$I\sim$0.8 mags. 
The $V-I$ and $R-I$ colors are significantly bluer than the observed early-time colors of SN 2002hh, however these colors are reasonably consistent with the reddened supernova light scattering from Galactic-type dust with a modest optical depth $A_V\sim 0.5$ \citep{2003AJ....126.1939S}.
In this scenario, the echo light passes through the same dust that is reddening the directly transmitted light from SN 2002hh and then passes through and is scattered off of dust lying nearby which also adds a small amount of reddening. 

No line-of-sight sources which might contaminate the lightcurve
and spectroscopy of SN 2002hh are seen in the shifted and combined 
sum of three 120 s R-band images (550745o, 550746o, and 550747o) taken 
on UT 2000 September 27 prior to outburst and shown in Figure 4. The 
seeing and depth of the combined image is very similar to our Gemini r' 
images and allows us to constrain the precursor to be fainter than R=23.0 
(or 7.5 magnitudes fainter than maximum light). To further clarify the
relative strength of the continuum and H$\alpha$ line at late times, the
unfluxed but sky-subtracted Gemini/GMOS spectrum for SN 2002hh is shown 
in Figure 5. The ratio of the peak H$\alpha$ to blue side continuum is 
5.0 and the ratio of H$\alpha$ to red side continuum is 2.9. These numbers 
are very similar to their day 44 counterparts of 4.75 and 2.7 
\citep{2006MNRAS.368.1169P}, respectively, and fully consistent with the 
light echo explanation for the late-time photometry and spectroscopy.

We note that the appearance of a resolved echo clearly indicates the
dominance of the echo light above the local background.
In an aperture surrounding the echo in F606W, the sky value is 2.4E-3 (c/s)/pix and in F658N, it is 7.4E-4 (c/s)/pix. This compares to peak values of the echo 
of 0.73 (c/s)/pix in F606W, and 9.2E-2 in F658N. In other words, the peak 
surface brightness of the echo is 300 times brighter than the background in 
F606W, and 125 times brighter in F658N.

While the bluer $V$-$I$ and $R$-$I$ colors are consistent with the scenario where the echo light is 
significantly less reddened than the direct SN light, the $B$-$V$ color measured by Sugerman 
et al. (2007, in preparation) is much too red. SN 2002hh is quite 
faint in the $B$ band and this measurement has a large uncertainty 
($>$0.5 mag) so the $B$-$V$ color could be bluer than reported here.
If the these colors are based solely on a less reddened path, then the 
$V$-$I$ and $R$-$I$ colors imply that the echo light passes through dust 
with an A$_V$ that is 2-2.5 mag less than the direct light from the SN. 
This scenario would serve to explain at least some of the excess late-time 
brightness of the SN 2002hh echo. For the time being, SN 2002hh is unique - it
appears to have the brightest echo yet observed around a SN, by a wide margin. 

\acknowledgments
We would like to thank the referee for many useful suggestions.
DLW was supported by the Natural Sciences and Engineering Research Council 
of Canada (NSERC). This project was also supported by the NSF/REU grant 
AST-0097694 and the Nantucket Maria Mitchell Association. We would like 
to thank Eric Feigelson at PSU for suggesting this as an interesting 
topic for investigation. Observations were obtained during programs, 
GN-2004B-C-3, GN-2004B-DD-6, GN-2005B-Q-54, and GN-2006A-Q-52 at the 
Gemini Observatory, which is operated by AURA under a cooperative 
agreement with the NSF on behalf of the Gemini partnership. We are 
indebted to the Gemini Observatory and its staff for helping to obtain 
these data and also for providing DD time for further observations of 
SN 2002hh. This research used the facilities of the Canadian Astronomy 
Data Centre operated by the National Research Council of Canada with 
the support of the Canadian Space Agency.

\appendix
\section{Tertiary Standards for NGC 6946}
The frequency with which supernovae have appeared in NGC 6946 during
the last century suggests that the definition of a local set of faint 
tertiary standards, suitable for use on large telescopes, would be 
a worthwhile undertaking. In Section 2, we described the determination of
the transformations to the $V$, $R$, and $I$ system of local photometric
standards list by \citet{2006MNRAS.368.1169P} in their Table 1.
We defined a set of fainter standards in the following way. Stars 
selected were present in our output photometry lists in all three filters 
($g'$, $r'$, and $i'$), at all 6 epochs, and for which the photometric
instrumental uncertainties were all less than or equal to 0.03 mag. 
There were 110 such stars. These are labelled 101-210 so that there is 
no confusion with the Pozzo et al. numbering scheme.

The instrumental $g'$, $r'$, and $i'$ magnitudes were transformed to 
the $VRI$ system determined by \citet{2006MNRAS.368.1169P} and the 
unweighted mean and standard deviation of the six points for each star 
are listed Table 2. The uncertainties don't formally include the 
uncertainties in the transformation - they are the standard deviations 
for the set of six transformed magnitudes in each filter.
The locations of the standard stars are plotted in Figure 6 which 
is an $r'$ image with an exposure of 60.0 s obtained with GMOS-N on 
2004 August 20. North is 10\degree CCW from vertical and east is 90 \degree
CW from north. The image scale is 0\farcs1454 per pixel and the height
and width of the image are 149 and 335 arcsec, respectively.

\clearpage

\begin{deluxetable}{lcccll}
\tablecaption{Half-intensity H$\alpha$ Wavelengths and Photometry for SN 2002hh}
\tablenum{1}
\tablehead{\colhead{Obs Date}&
           \colhead{Age}&
           \colhead{50\% H$\alpha$ $\lambda_{asc}$}&
           \colhead{50\% H$\alpha$ $\lambda_{des}$} & \colhead{R$^a$}& \colhead{I$^a$}\\
           \colhead{(UT)}&
           \colhead{(days)}&
           \colhead{(\AA)}&
           \colhead{(\AA)}&
            \colhead{(mag)}&
             \colhead{(mag)}
           }
\startdata
2004 August 20&661&6499.9&6618.8&19.87   &  18.98 \\
2004 October 8&710&6497.2&6620.8&19.91  &   19.01\\
2004 December 11&774&6499.6&6625.5&19.97    & 19.13 \\
2005 August 4&1010&6498.9&6619.6&20.15   &  19.35 \\
2005 October 4&1071&6500.2&6623.2&20.15   &  19.40\\
2006 July 18&1358&6497.9&6621.1&\nodata&19.60\\
\enddata
\tablenotetext{a}{The $R$ and $I$ magnitudes were converted from the 
observed $g'$, $r'$, and $i'$ instrumental magnitudes. See text.}
\end{deluxetable}

\clearpage

\begin{deluxetable}{ccccccc}
\tablecaption{Tertiary $VRI$ Photometric Standards for NGC 6946}
\tablenum{2}
\tablehead{\colhead{Star Nbr.} &  \colhead{V} & \colhead{$\sigma_V$} & \colhead{R} & \colhead{$\sigma_R$} & \colhead{I} & \colhead{$\sigma_I$} \\ 
\colhead{} &  \colhead{(mag)} & \colhead{(mag)} & \colhead{(mag)} & \colhead{(mag)} & \colhead{(mag)} & \colhead{(mag)} } 

\startdata
101& 20.31 & 0.03 & 19.35 & 0.04 & 18.42 & 0.05 \\
102& 18.71 & 0.02 & 18.07 & 0.02 & 17.46 & 0.03 \\
103& 20.14 & 0.01 & 19.23 & 0.02 & 17.71 & 0.02 \\
104& 20.89 & 0.03 & 20.30 & 0.07 & 19.65 & 0.10 \\
105& 19.95 & 0.02 & 19.41 & 0.02 & 18.84 & 0.04 \\
106& 17.82 & 0.01 & 17.37 & 0.01 & 16.88 & 0.02 \\
107& 20.54 & 0.03 & 20.02 & 0.05 & 19.46 & 0.08 \\
108& 20.11 & 0.02 & 19.56 & 0.01 & 19.03 & 0.04 \\
109& 18.15 & 0.01 & 17.42 & 0.01 & 16.80 & 0.01 \\
110& 19.80 & 0.01 & 19.07 & 0.00 & 18.40 & 0.01 \\
111& 17.38 & 0.01 & 16.92 & 0.01 & 16.43 & 0.01 \\
112& 20.74 & 0.02 & 20.39 & 0.02 & 20.02 & 0.03 \\
113& 20.53 & 0.01 & 19.77 & 0.01 & 19.10 & 0.02 \\
114& 19.86 & 0.01 & 19.09 & 0.02 & 18.40 & 0.02 \\
115& 19.86 & 0.01 & 18.83 & 0.01 & 17.76 & 0.01 \\
116& 20.29 & 0.00 & 19.22 & 0.01 & 18.09 & 0.01 \\
117& 20.53 & 0.02 & 19.59 & 0.01 & 18.78 & 0.02 \\
118& 19.15 & 0.01 & 18.63 & 0.01 & 18.10 & 0.01 \\
119& 20.29 & 0.01 & 19.63 & 0.01 & 19.06 & 0.03 \\
120& 17.43 & 0.01 & 16.92 & 0.01 & 16.41 & 0.01 \\
121& 19.86 & 0.01 & 19.14 & 0.00 & 18.51 & 0.01 \\
122& 18.69 & 0.01 & 18.01 & 0.01 & 17.40 & 0.01 \\
123& 20.48 & 0.01 & 19.80 & 0.00 & 19.19 & 0.01 \\
124& 18.44 & 0.01 & 17.48 & 0.01 & 16.58 & 0.01 \\
125& 17.57 & 0.01 & 17.01 & 0.01 & 16.47 & 0.01 \\
126& 19.00 & 0.01 & 18.30 & 0.01 & 17.69 & 0.01 \\
127& 19.43 & 0.01 & 18.69 & 0.00 & 18.03 & 0.01 \\
128& 18.67 & 0.01 & 18.02 & 0.01 & 17.43 & 0.01 \\
129& 19.50 & 0.01 & 18.62 & 0.01 & 17.87 & 0.01 \\
130& 18.96 & 0.04 & 18.41 & 0.04 & 17.87 & 0.02 \\
131& 19.01 & 0.01 & 18.40 & 0.01 & 17.85 & 0.01 \\
132& 20.31 & 0.02 & 19.27 & 0.01 & 17.87 & 0.01 \\
133& 18.71 & 0.01 & 18.23 & 0.01 & 17.76 & 0.01 \\
134& 18.65 & 0.01 & 17.97 & 0.00 & 17.35 & 0.01 \\
135& 20.77 & 0.04 & 19.93 & 0.03 & 19.22 & 0.03 \\
136& 20.98 & 0.01 & 20.33 & 0.01 & 19.72 & 0.02 \\
137& 20.10 & 0.01 & 19.38 & 0.01 & 18.77 & 0.02 \\
138& 19.75 & 0.01 & 19.20 & 0.01 & 18.65 & 0.01 \\
139& 17.71 & 0.01 & 17.15 & 0.01 & 16.62 & 0.01 \\
140& 18.68 & 0.01 & 18.19 & 0.00 & 17.70 & 0.01 \\
141& 20.21 & 0.02 & 19.18 & 0.01 & 18.19 & 0.01 \\
142& 19.39 & 0.00 & 18.76 & 0.01 & 18.20 & 0.00 \\
143& 19.25 & 0.01 & 18.67 & 0.01 & 18.09 & 0.01 \\
144& 19.17 & 0.01 & 18.62 & 0.00 & 18.10 & 0.01 \\
145& 17.67 & 0.01 & 17.16 & 0.01 & 16.67 & 0.01 \\
146& 18.89 & 0.01 & 18.31 & 0.01 & 17.75 & 0.01 \\
147& 18.31 & 0.01 & 17.51 & 0.01 & 16.83 & 0.01 \\
148 & 20.66 & 0.01 & 20.08 & 0.01 & 19.53 & 0.01 \\
149 & 19.48 & 0.00 & 18.80 & 0.00 & 18.19 & 0.00 \\
150 & 17.53 & 0.00 & 16.92 & 0.01 & 16.35 & 0.01 \\
151 & 19.87 & 0.01 & 19.25 & 0.01 & 18.69 & 0.00 \\
152 & 20.62 & 0.04 & 19.70 & 0.02 & 18.84 & 0.02 \\
153 & 17.44 & 0.01 & 16.77 & 0.00 & 16.12 & 0.01 \\
154 & 20.94 & 0.03 & 20.08 & 0.04 & 19.29 & 0.05 \\
155 & 20.70 & 0.01 & 19.59 & 0.01 & 18.54 & 0.01 \\
156 & 21.03 & 0.01 & 20.43 & 0.03 & 20.32 & 0.03 \\
157 & 19.66 & 0.02 & 18.77 & 0.01 & 18.01 & 0.01 \\
158 & 19.54 & 0.02 & 18.66 & 0.00 & 16.87 & 0.01 \\
159 & 18.07 & 0.01 & 17.43 & 0.00 & 16.85 & 0.00 \\
160 & 18.46 & 0.01 & 17.89 & 0.01 & 17.33 & 0.01 \\
161 & 19.98 & 0.01 & 19.27 & 0.01 & 18.64 & 0.01 \\
162 & 19.81 & 0.01 & 18.74 & 0.01 & 17.38 & 0.00 \\
163 & 19.60 & 0.02 & 18.88 & 0.01 & 18.21 & 0.01 \\
164 & 17.87 & 0.01 & 17.14 & 0.01 & 16.49 & 0.01 \\
165 & 20.83 & 0.01 & 20.18 & 0.02 & 19.58 & 0.02 \\
166 & 20.54 & 0.01 & 19.84 & 0.00 & 19.20 & 0.01 \\
167 & 20.69 & 0.02 & 19.72 & 0.01 & 18.80 & 0.01 \\
168 & 20.67 & 0.03 & 19.59 & 0.01 & 18.63 & 0.01 \\
169 & 18.61 & 0.00 & 18.08 & 0.00 & 17.54 & 0.01 \\
170 & 19.11 & 0.01 & 18.48 & 0.01 & 17.86 & 0.01 \\
171 & 20.67 & 0.02 & 20.30 & 0.02 & 19.82 & 0.03 \\
172 & 17.18 & 0.01 & 16.43 & 0.01 & 15.80 & 0.01 \\
173 & 18.76 & 0.01 & 18.10 & 0.01 & 17.47 & 0.01 \\
174 & 17.71 & 0.01 & 17.02 & 0.00 & 16.34 & 0.01 \\
175 & 20.37 & 0.03 & 19.51 & 0.01 & 18.76 & 0.03 \\
176 & 21.36 & 0.03 & 20.57 & 0.01 & 19.88 & 0.02 \\
177 & 17.81 & 0.01 & 17.29 & 0.00 & 16.76 & 0.00 \\
178 & 18.46 & 0.01 & 17.83 & 0.01 & 17.23 & 0.01 \\
179 & 18.84 & 0.01 & 18.16 & 0.01 & 17.52 & 0.01 \\
180 & 19.65 & 0.01 & 18.87 & 0.02 & 18.25 & 0.01 \\
181 & 19.62 & 0.01 & 18.99 & 0.01 & 18.39 & 0.02 \\
182 & 18.96 & 0.01 & 18.48 & 0.01 & 17.97 & 0.01 \\
183 & 21.13 & 0.03 & 20.32 & 0.02 & 19.64 & 0.01 \\
184 & 20.54 & 0.04 & 19.61 & 0.03 & 18.81 & 0.02 \\
185 & 20.66 & 0.02 & 20.30 & 0.03 & 19.81 & 0.04 \\
186 & 20.29 & 0.02 & 19.31 & 0.02 & 17.78 & 0.02 \\
187 & 17.98 & 0.01 & 17.09 & 0.01 & 16.30 & 0.01 \\
188 & 19.80 & 0.02 & 18.76 & 0.01 & 17.54 & 0.01 \\
189 & 16.91 & 0.01 & 16.34 & 0.01 & 15.76 & 0.01 \\
190 & 19.94 & 0.02 & 18.85 & 0.01 & 17.75 & 0.02 \\
191 & 19.49 & 0.01 & 18.75 & 0.01 & 18.10 & 0.01 \\
192 & 20.58 & 0.03 & 19.59 & 0.02 & 18.29 & 0.02 \\
193 & 19.49 & 0.01 & 18.84 & 0.01 & 18.22 & 0.01 \\
194 & 20.24 & 0.02 & 19.48 & 0.01 & 18.78 & 0.02 \\
195 & 19.05 & 0.01 & 18.31 & 0.00 & 17.63 & 0.01 \\
196 & 17.56 & 0.01 & 16.89 & 0.01 & 16.28 & 0.02 \\
197 & 17.99 & 0.01 & 17.41 & 0.01 & 16.89 & 0.02 \\
198 & 20.17 & 0.01 & 19.53 & 0.00 & 18.92 & 0.02 \\
199 & 19.11 & 0.01 & 18.29 & 0.01 & 17.57 & 0.02 \\
200 & 18.80 & 0.01 & 18.27 & 0.01 & 17.71 & 0.02 \\
201 & 20.68 & 0.02 & 20.08 & 0.02 & 19.47 & 0.06 \\
202 & 20.70 & 0.02 & 20.18 & 0.01 & 19.64 & 0.02 \\
203 & 20.20 & 0.01 & 19.40 & 0.01 & 18.70 & 0.02 \\
204 & 19.40 & 0.01 & 18.50 & 0.01 & 17.70 & 0.02 \\
205 & 21.05 & 0.05 & 20.44 & 0.05 & 20.03 & 0.04 \\
206 & 16.98 & 0.01 & 16.46 & 0.01 & 15.94 & 0.01 \\
207 & 19.05 & 0.01 & 18.40 & 0.01 & 17.77 & 0.02 \\
208 & 18.91 & 0.01 & 18.34 & 0.01 & 17.76 & 0.01 \\
209 & 21.21 & 0.04 & 20.52 & 0.03 & 19.86 & 0.02 \\
210 & 20.32 & 0.09 & 19.36 & 0.07 & 18.59 & 0.06 \\
\enddata
\end{deluxetable}

\clearpage

\begin{figure}
\figurenum{1}
\begin{center}
\includegraphics[width=5in,angle=0]{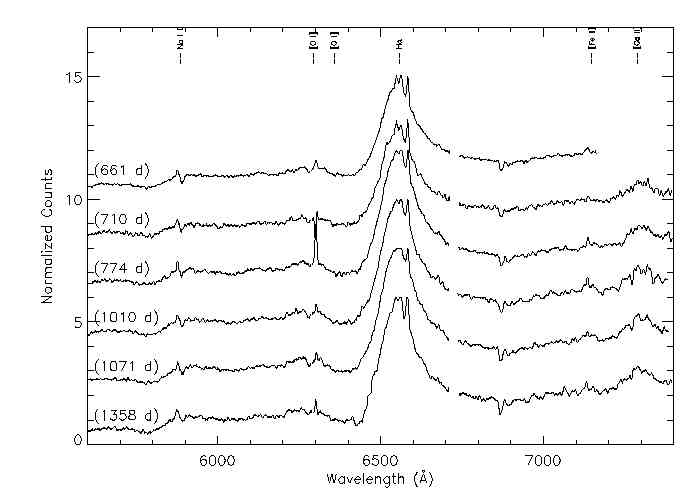}
\end{center}
\caption{New Gemini/GMOS spectra of SN 2002hh from day 661 to day 1358. 
For display purposes, the narrow emission lines from NGC 6946 have been 
truncated.
}
\end{figure}

\clearpage

\begin{figure}
\figurenum{2}
\begin{center}
\includegraphics[width=5in,angle=0]{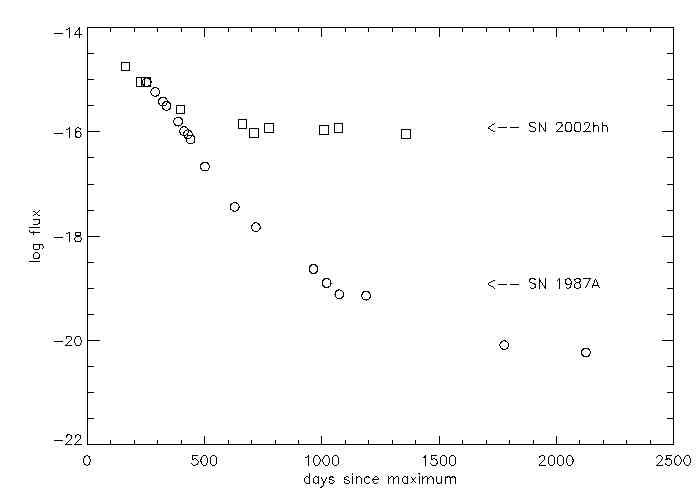}
\end{center}
\caption{Measures of the peak counts in H$\alpha$ in SN 1987A (circles) 
and SN 2002hh (squares) plotted against time since maximum light.}
\end{figure}

\clearpage

\begin{figure}
\figurenum{3}
\begin{center}
\includegraphics[width=5in,angle=0]{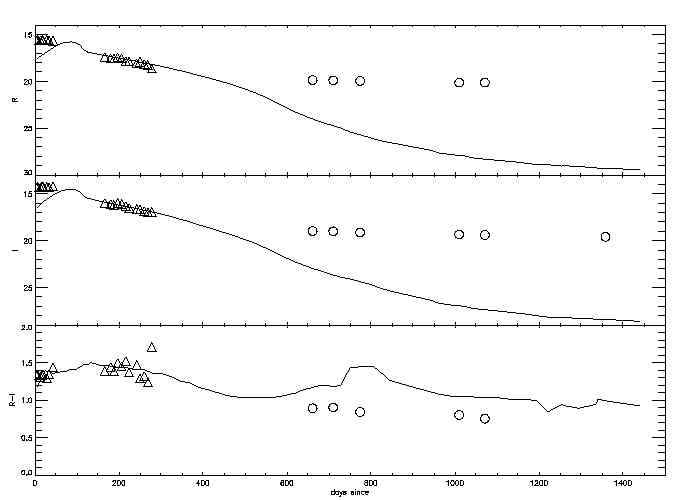}
\end{center}
\caption{Lightcurve of SN 2002hh plotted against a scaled lightcurve 
from SN 1987A (solid line) \protect\citep{1990AJ.....99.1146H, 1991PASP..103..958W} . 
The upper two panels show the observed $R$ and $I$ photmetry while the 
lower panel show the $R$-$I$ colors. The SN 1987A $R$-$I$ colors have 
been adjusted asssuming A$_V$=5 mag and R$_V$=3.1. The SN 2002hh photometry 
is from \protect\citet{2006MNRAS.368.1169P} (triangles) and from this study (circles).
}
\end{figure}

\begin{figure}
\figurenum{4}
\begin{center}
\includegraphics[width=5in,angle=0]{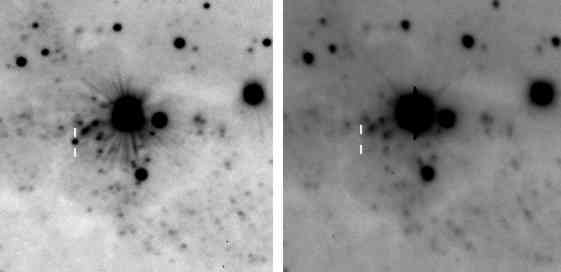}
\end{center}
\caption{The Gemini/GMOS r' image taken on 2004 August 20 (left) is shown along with the R-band image taken with the pre-outburst CFHT CFH12K image on 2000 
September 27 (right). The position of SN 2002hh is marked on both images. 
The supernova is R$\sim$19.9 mag on 2004 August 20 and there is an upper limit 
of $\sim$23.0 mag for any confusing source at the position of SN 2002hh on 
2000 September 27. The field is 40\arcsec x 40\arcsec. North is up and east 
is to the left.} 
\end{figure}

\begin{figure}
\figurenum{5}
\begin{center}
\includegraphics[width=5in,angle=0]{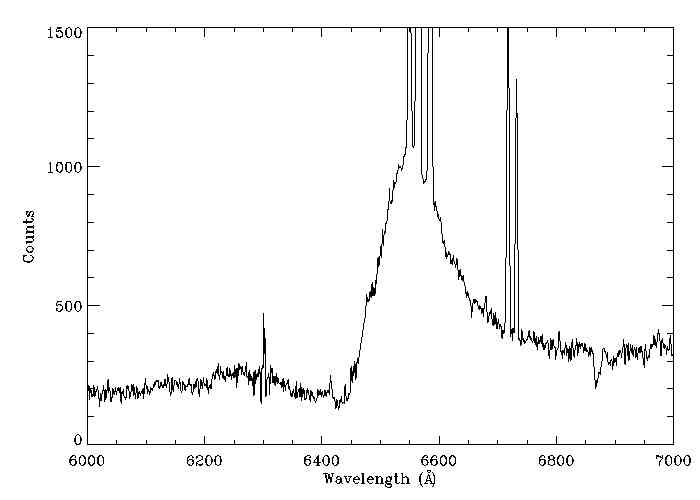}
\end{center}
\caption{Gemini/GMOS spectrum of SN 2002hh taken on 2006 July 18. The plotted 
spectrum is the average of three individual spectra taken on the same date. 
The ratios of the peak line emission to the continuum on either side of the 
H$\alpha$ line are very similar to those seen in the spectrum taken of 
SN 2002hh on day 44.} 
\end{figure}

\clearpage

\begin{figure}
\figurenum{6}
\begin{center}
\includegraphics[width=3in,angle=0]{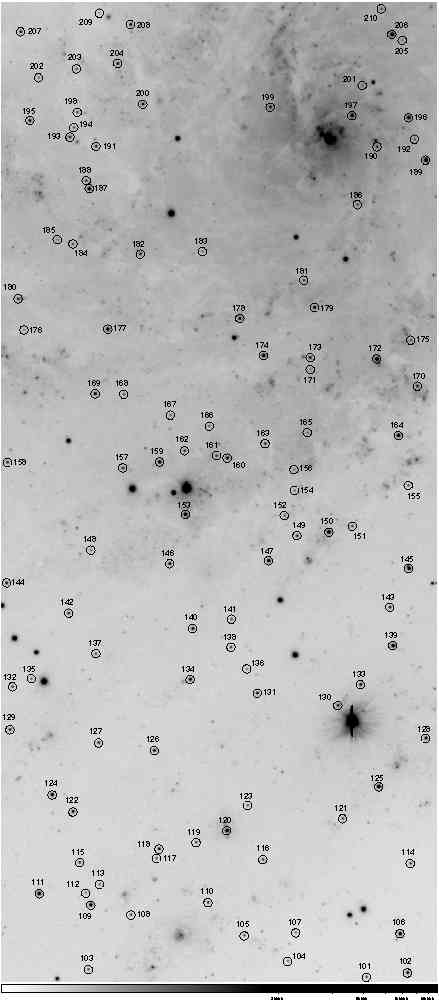}
\end{center}
\caption{Stars included in Table 2 as tertiary standards.}
\end{figure}


\begin{thebibliography}{27}
\expandafter\ifx\csname natexlab\endcsname\relax\def\natexlab#1{#1}\fi
\expandafter\ifx\csname href\endcsname\relax
  \def\href#1#2{}\fi
\expandafter\ifx\csname urllinklabel\endcsname\relax
  \def\urllinklabel{[LINK]}\fi
\expandafter\ifx\csname adsurllinklabel\endcsname\relax
  \def\adsurllinklabel{[ADS]}\fi

\bibitem[{{Barlow et al.}(2005)}]{2005ApJ...627L.113B}
{Barlow et al.}, M.~J. 2005, \apjl, 627, L113


\bibitem[{{Caldwell et al.}(1993)}]{1993MNRAS.262..313C}
{Caldwell et al.}, J.~A.~R. 1993, \mnras, 262, 313


\bibitem[{{Danziger} {et~al.}(1991){Danziger}, {Lucy}, {Bouchet}, \&
  {Gouiffes}}]{1991supe.conf...69D}
{Danziger}, I.~J., {Lucy}, L.~B., {Bouchet}, P., \& {Gouiffes}, C. 1991, in
  Supernovae. The Tenth Santa Cruz Workshop in Astronomy and Astrophysics,
  (Springer-Verlag, New York), ed. S.~E. {Woosley}, p. 69


\bibitem[{{Duschinger} {et~al.}(1995){Duschinger}, {Puls}, {Branch},
  {Hoeflich}, \& {Gabler}}]{1995A&A...297..802D}
{Duschinger}, M., {Puls}, J., {Branch}, D., {Hoeflich}, P., \& {Gabler}, A.
  1995, \aap, 297, 802


\bibitem[{{Elmhamdi et al.}(2003)}]{2003MNRAS.338..939E}
{Elmhamdi et al.}, A. 2003, \mnras, 338, 939


\bibitem[{{Filippenko} {et~al.}(2002){Filippenko}, {Foley}, \&
  {Swift}}]{2002IAUC.8007....2F}
{Filippenko}, A.~V., {Foley}, R.~J., \& {Swift}, B. 2002, \iaucirc, 8007, 2


\bibitem[{{Gerardy} {et~al.}(2000){Gerardy}, {Fesen}, {H{\"o}flich}, \&
  {Wheeler}}]{2000AJ....119.2968G}
{Gerardy}, C.~L., {Fesen}, R.~A., {H{\"o}flich}, P., \& {Wheeler}, J.~C. 2000,
  \aj, 119, 2968


\bibitem[{{Hamuy} \& {Suntzeff}(1990)}]{1990AJ.....99.1146H}
{Hamuy}, M. \& {Suntzeff}, N.~B. 1990, \aj, 99, 1146


\bibitem[{{Li}(2002)}]{2002IAUC.8005....1L}
{Li}, W. 2002, \iaucirc, 8005, 1


\bibitem[{{Lucy} {et~al.}(1989){Lucy}, {Danziger}, {Gouiffes}, \&
  {Bouchet}}]{1989LNP...350..164L}
{Lucy}, L.~B., {Danziger}, I.~J., {Gouiffes}, C., \& {Bouchet}, P. 1989,
  Structure and Dynamics of the Interstellar Medium, IAU Colloq.~120:, 350, 164


\bibitem[{{Lucy} {et~al.}(1991){Lucy}, {Danziger}, {Gouiffes}, \&
  {Bouchet}}]{1991supe.conf...82L}
{Lucy}, L.~B., {Danziger}, I.~J., {Gouiffes}, C., \& {Bouchet}, P. 1991, in
  Supernovae. The Tenth Santa Cruz Workshop in Astronomy and Astrophysics,
  (Springer-Verlag, New York), ed. S.~E. {Woosley}, 82


\bibitem[{{Mattila} \& {Meikle}(2001)}]{2001MNRAS.324..325M}
{Mattila}, S. \& {Meikle}, W.~P.~S. 2001, \mnras, 324, 325


\bibitem[{{McGough} {et~al.}(2005){McGough}, {Clayton}, {Gordon}, \&
  {Wolff}}]{2005ApJ...624..118M}
{McGough}, C., {Clayton}, G.~C., {Gordon}, K.~D., \& {Wolff}, M.~J. 2005, \apj,
  624, 118


\bibitem[{{Meikle} {et~al.}(2002){Meikle}, {Mattila}, {Smartt}, {MacDonald},
  {Clewley}, \& {Dalton}}]{2002IAUC.8024....1M}
{Meikle}, P., {Mattila}, S., {Smartt}, S., {MacDonald}, E., {Clewley}, L., \&
  {Dalton}, G. 2002, \iaucirc, 8024, 1


\bibitem[{{Meikle} {et~al.}(2006){Meikle}, {Mattila}, {Gerardy}, {Kotak},
  {Pozzo}, {van Dyk}, {Farrah}, {Fesen}, {Filippenko}, {Fransson}, {Lundqvist},
  {Sollerman}, \& {Wheeler}}]{2006ApJ...649..332M}
{Meikle}, W.~P.~S., {Mattila}, S., {Gerardy}, C.~L., {Kotak}, R., {Pozzo}, M.,
  {van Dyk}, S.~D., {Farrah}, D., {Fesen}, R.~A., {Filippenko}, A.~V.,
  {Fransson}, C., {Lundqvist}, P., {Sollerman}, J., \& {Wheeler}, J.~C. 2006,
  \apj, 649, 332


\bibitem[{{Patat}(2005)}]{2005MNRAS.357.1161P}
{Patat}, F. 2005, \mnras, 357, 1161


\bibitem[{{Patat} {et~al.}(2006){Patat}, {Benetti}, {Cappellaro}, \&
  {Turatto}}]{2006MNRAS.369.1949P}
{Patat}, F., {Benetti}, S., {Cappellaro}, E., \& {Turatto}, M. 2006, \mnras,
  369, 1949


\bibitem[{{Phillips} \& {Williams}(1991)}]{1991supe.conf...36P}
{Phillips}, M.~M. \& {Williams}, R.~E. 1991, in Supernovae. The Tenth Santa
  Cruz Workshop in Astronomy and Astrophysics, (Springer-Verlag, New York), ed.
  S.~E. {Woosley}, p. 36


\bibitem[{{Pozzo} {et~al.}(2004){Pozzo}, {Meikle}, {Fassia}, {Geballe},
  {Lundqvist}, {Chugai}, \& {Sollerman}}]{2004MNRAS.352..457P}
{Pozzo}, M., {Meikle}, W.~P.~S., {Fassia}, A., {Geballe}, T., {Lundqvist}, P.,
  {Chugai}, N.~N., \& {Sollerman}, J. 2004, \mnras, 352, 457


\bibitem[{{Pozzo et al.}(2006)}]{2006MNRAS.368.1169P}
{Pozzo et al.}, M. 2006, \mnras, 368, 1169


\bibitem[{{Sahu} {et~al.}(2006){Sahu}, {Anupama}, {Srividya}, \&
  {Muneer}}]{2006astro.ph..8432S}
{Sahu}, D.~K., {Anupama}, G.~C., {Srividya}, S., \& {Muneer}, S. 2006,
  astro-ph/0608432


\bibitem[{{Sirianni} {et~al.}(2005){Sirianni}, {Jee}, {Ben{\'{\i}}tez},
  {Blakeslee}, {Martel}, {Meurer}, {Clampin}, {De Marchi}, {Ford}, {Gilliland},
  {Hartig}, {Illingworth}, {Mack}, \& {McCann}}]{2005PASP..117.1049S}
{Sirianni}, M., {Jee}, M.~J., {Ben{\'{\i}}tez}, N., {Blakeslee}, J.~P.,
  {Martel}, A.~R., {Meurer}, G., {Clampin}, M., {De Marchi}, G., {Ford}, H.~C.,
  {Gilliland}, R., {Hartig}, G.~F., {Illingworth}, G.~D., {Mack}, J., \&
  {McCann}, W.~J. 2005, \pasp, 117, 1049


\bibitem[{{Sugerman}(2003)}]{2003AJ....126.1939S}
{Sugerman}, B.~E.~K. 2003, \aj, 126, 1939


\bibitem[{{Sugerman et al.}(2006)}]{2006Sci...313..196S}
{Sugerman et al.}, B.~E.~K. 2006, Science, 313, 196


\bibitem[{{Van Dyk} {et~al.}(2006){Van Dyk}, {Li}, \&
  {Filippenko}}]{2006PASP..118..351V}
{Van Dyk}, S.~D., {Li}, W., \& {Filippenko}, A.~V. 2006, \pasp, 118, 351


\bibitem[{{Walker} \& {Suntzeff}(1991)}]{1991PASP..103..958W}
{Walker}, A.~R. \& {Suntzeff}, N.~B. 1991, \pasp, 103, 958


\bibitem[{{Wang}(2005)}]{2005ApJ...635L..33W}
{Wang}, L. 2005, \apjl, 635, L33


\end{thebibliography}
\end{document}